# Local Betweenness Centrality Analysis of 30 European Cities

**Yamaoka Kaoru**
yamaoka.kaoru@nikken.jp
NIKKEN SEKKEI Ltd.
Tokyo 102-8117, Japan

**Kumakoshi Yusuke**
ykuma@cd.t.u-tokyo.ac.jp
Research Center for Advanced Science and Technology
The University of Tokyo
Tokyo 153-8904, Japan

**Yoshimura Yuji**
yyyoshimura@cd.t.u-tokyo.ac.jp
Research Center for Advanced Science and Technology
The University of Tokyo
Tokyo 153-8904, Japan

Abstract†† Urban morphology and socioeconomic aspects of cities have been explored by analysing urban street network. To analyse the network, several variations of the centrality indices are often used. However, its nature has not yet been widely studied, thus leading to an absence of robust visualisation method of urban road network characteristics. To fill this gap, we propose to use a set of local betweenness centrality and a new simple and robust visualisation method. By analysing 30 European cities, we found that our method illustrates common structures of the cities: road segments important for long-distance transportations are concentrated along larger streets while those for short range transportations form clusters around CBD, historical, or residential districts. Quantitative analysis has corroborated these findings. Our findings are useful for urban planners and decision-makers to understand the current situation of the city and make informed decisions.

# 1 Introduction

This paper discusses the methodology to distinguish the pedestrian-oriented street from others using the local betweenness centrality indicator. The street network is a topological network deriving from urban road geometry and it is a fundamental basis of our city. To analyse them, we propose using local betweenness indicator. Betweenness centrality has been widely used for urban analysis since the 1970's and had helped to uncover various aspects of cities. However, the whole nature of the indicator is not yet clear. For example, this indicator does not give importance both for motorized transportations and pedestrians, or whether structures of those importance are common among cities. In this paper, we attempt to uncover these aspects, focusing on thirty European cities and its analysis.

Attempts to understand built environments through their geometric patterns have been made in the realm of urban morphology studies. Some of them are strongly related to road networks or road environments: visibility analysis (Chamberlain & Meitner, 2013; O'Sullivan & Turner, 2001) for finer scale analysis, Axial line analysis (Bafna, 2003; Hillier, 1996; Penn, 2003) and road network analysis (Peponis et al., 2008) for larger scale analysis.

One of the most common approaches to extract important features from road network structures is to regard them as a network, and then perform various network analysis (Barabási & Pósfai, 2016). Centrality is an indicator for the importance of each network element in a network and has several variants (Porta et al., 2006, 2009). The centrality analysis can be used to determine important road segments on road networks, in terms of human activities which take place on the road networks.

Several centrality measures have been used in analysing the road network, such as betweenness centrality (Barabási & Pósfai, 2016; Linton C. Freeman, 1978), closeness centrality (Linton C. Freeman, 1978; Nieminen, 1974), straightness centrality (Crucitti et al., 2006) or combination of them (Crucitti et al., 2006; Porta et al., 2006, 2009; Strano et al., 2012). Betweenness centrality of a network element is a number of all shortest paths in the network that go through the element. Closeness centrality indicates if other nodes are accessible from a node in shorter distances. Straightness centrality is a similar measure as closeness which uses the distance along the network divided by Euclidean distance instead of distance. See Barabási & Pósfai (2016) for the formal definitions of these measures.

Within these measures we chose betweenness centrality for the analysis of this paper because betweenness centrality takes not only the origin and the destination of paths but also the route between them. This is particularly important when we regard centrality as a simplified model of human activities because it can express detouring or stopping-by behaviours.

However, naïve betweenness centrality analysis has several problems. First, betweenness centrality analysis has the "edge-effect" (quantitatively examined in Gil, 2017; Okabe & Sugihara, 2012, p.41; Ratti, 2004), which means that estimated betweenness of a network is sensitive to the edge at which the network is bordered. With this effect, features near the centre of the region of interest (ROI) tend to have higher degree of betweenness centrality compared with those close to the edge. Second, naïve betweenness centrality cannot represent multiple aspects that can be seen in actual cities (Porta et al., 2006). For example, a small road segment in a residential area might be important for pedestrians, but almost negligible for motorised transportation. In this sense, the road segment should have at least two kinds of "importance": one for pedestrians and the other for motorised transportation. Nevertheless, naïve betweenness centrality analysis assigns only 1-dimensional value for it.

To avoid these problems, we propose using a set of local betweenness centrality (Yoshimura et al., 2020). This centrality measure takes paths within a threshold radius into account instead of considering all possible paths in the city. With this measure, one can spot features important for movements within an arbitrary threshold radius, for example 500 metres or 5 kilometres.

The idea of varying radius for centrality computation appears in some recent research. Porta et al. uses 800, 1600, 2400 metres for localised closeness centrality (Porta et al., 2012). Pont et al. (2019) uses straightness betweenness radius varying from 500 meters to 5000 meters with 500 meters while Yoshimura et al. (2020) uses betweenness centrality with radius varying from 300 to 5000 metres with 100 meters step (Yoshimura et al., 2020).

Multidimensional data can be obtained from a road network by varying the threshold radius, although there is no obvious interpretation method for the data. Yoshimura et al. (2020) and Pont et al. (2019) have used clustering algorithms for this purpose.

However, we found that clustering methods are not necessarily universal nor robust.

First, the results of clustering methods are affected by the choice of the number of clusters. Second and more importantly, the characteristics of clusters are not guaranteed to be consistent among cities. To avoid these shortcomings, we propose a new simple and robust visualisation method. The method can show not only the importance of a road segment as conventional betweenness centrality visualisation does, but also the characteristic of the road segment.

Our contributions are the followings: First, our method of analysis is visually easy to interpret. Second, we developed a new simple and robust method for visualizing datasets coming from different cities. Finally, we applied the method to multiple cities in Europe and found common tendencies in larger scale structure and some differences in smaller scale structure.

This paper is organized in the following manner: Section 2 describes the methods we used. Section 3 describes experiments we have conducted and its results. In section 4 we discuss on our result and future work. Section 5 describes conclusion.

## 2 Methodology

### 2.1 Datasets and Study Area

We used OSMnx (Boeing, 2017) to obtain road centreline data from the OpenStreetMap (OSM). The OSM has large data coverage and its data is open; it was therefore suitable for our research. However, the data derived from the OSM can have imperfections. For instance, areas with a small population might be left intact or with different cultural backgrounds might have different input tendencies. To avoid these problems, we chose the 10 largest populated cities from France, Spain and Italy for analysis, resulting in 30 road network data.
We used a 5000m buffer to download the road network in order to mitigate edge effects. The buffer regions were trimmed after betweenness computation and not included in the analysis.

Raw data of road networks downloaded from the OSM contains trivial roads such as park trails. These roads are usually narrow and does not have any building alongside thus not suitable to be an origin or a destination in betweenness analysis. To avoid including these, we used roads that can be used by bicycles for analysis. With this condition, we were able to exclude small paths as shown in Fig. 1.. Then, parallel road segments are unified using ArcGIS (as seen in Long & Liu, 2017). Vertices within 0.1 meters are unified to avoid positional error.

Finally, redundant road segments which share the same origin and destination nodes with other segments are merged into one road segment. In this merging process, a road segment with the shortest length was taken. This is because betweenness centrality analysis uses the shortest path for analysis and therefore longer segments will never be taken into account in the analysis.

Also data of Barcelona city was additionally used as a reference input data to check if algorithms are working properly. Because the street network dataset came from CartoBCN, the official website of Barcelona city council, we identify that it should contain all proper road segment information and thus can be used as a reference data.

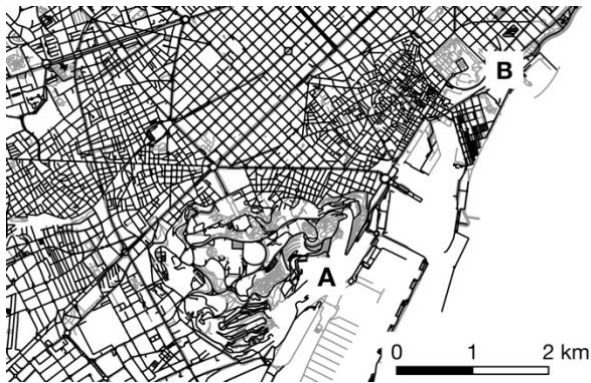

**Fig. 1.** Roads that can be used by bicycles (black) and pedestrians (grey). Latter includes park trails. These road segments are typically seen in large parks such as Barcelona Botanical Garden (A) or Barcelona Zoo (B).

## 2.2 Proposed indicators: A Set of Local Betweenness Centralities

We propose to compute a set of local betweenness centralities for the road networks. The indicator is defined as follows.

### Betweenness Centrality of a road network

Betweenness centrality, in its conventional usage, is a centrality measure for a graph which gives higher value for network features that are included in more of the all shortest paths in the graph. To compute betweenness centrality of a road network, it needs to be represented as a graph. In most cases vertices and edges corresponds to crossings and road segments, respectively.

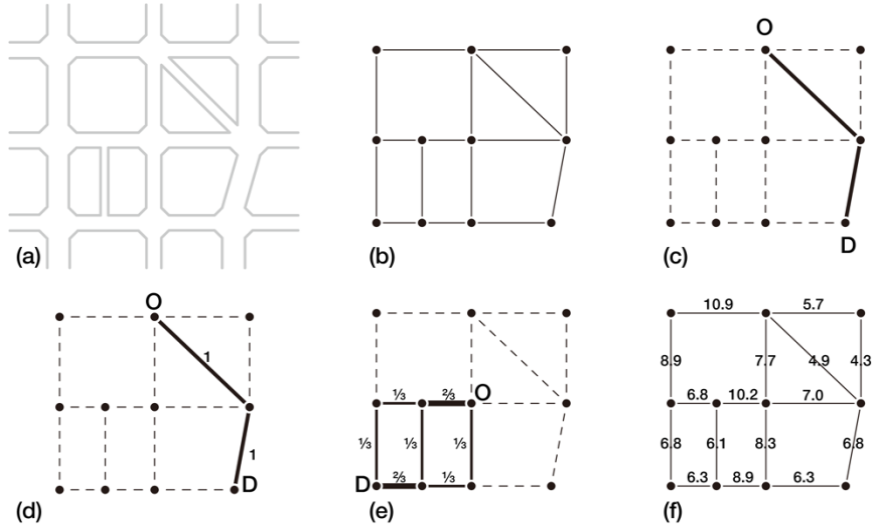

**Fig. 2.** Steps to compute edge betweenness centrality for a road network. (a) road network, (b) road network represented as a graph, (c) a pair of origin and destination and shortest paths between them, (d)(e) contribution of each O/D pair and corresponding shortest path(s) to overall betweenness centrality. (f) Betweenness centrality computed by summing up all of the O/D pair in the network.

The conventional betweenness centrality for a vertex or an edge is defined as follows:

$$g(v) = \sum_{s \neq v \neq t} \frac{\sigma_{st}(v)}{\sigma_{st}}$$

where $g(v)$ is the betweenness centrality, $\sigma_{st}$ is the number of all shortest paths, and $\sigma_{st}(v)$ is the number of all shortest paths that go through vertex $v$. Same can be applied for edges.

**Local Betweenness Centrality**

Local betweenness centrality is a variant of the betweenness centrality in which only shortest paths whose length are shorter than a threshold are taken into consideration. By limiting the path length, it can extract more local or smaller scale features of the network compared to the original betweenness centrality measure. The measure was first introduced by Borgatti and Everett (2006). The local betweenness centrality can be noted as:

$$g(v, \theta) = \sum_{s \neq v \neq t} \frac{\sigma_{st}(v, \theta)}{\sigma_{st}}$$

where $\sigma_{st}(v, \theta)$ is the number of all shortest paths that go through vertex v and has length shorter than the threshold $\theta$. Other notations are the same as the conventional betweenness centrality. Same can be applied for edges.

All centrality values are finally normalised to range [0, 1] along all vertices as recent research does (Leydesdorff, 2007). This is to understand relative importance within the ROI.

Both can be computed efficiently using Brandes' algorithm (Brandes, 2001). In Brandes' algorithm, Dijkstra's algorithm's (W Dijkstra, 1959) inner state is used to compute a set of shortest paths from a vertex in the graph. Since Dijkstra's algorithm is a best-first search algorithm, it is guaranteed to find all shortest paths shorter than current path length even if aborted on the way. For this reason, one can abort Dijkstra's algorithm if the distance from origin to algorithm's "current" vertices is larger than a threshold. After aborting the algorithm, paths longer than threshold are trimmed, and the resulting inner state is accumulated to betweenness centrality values. One can efficiently compute betweenness for all radii by repeating this trimming and accumulating process instead of running Dijkstra's algorithm for each radius.

**A Set of Local Betweenness Centralities**

Results of the local betweenness centrality is heavily affected by its radius as shown in latter sections. Therefore we computed local betweenness centrality values for 300 to 5000 meters radius by 100m steps for each city. The former radius corresponds to short-range transportation such as pedestrian movement and the latter corresponds to long-range, or motorised transportation.

## 2.3 Computation

We used Julia language and LightGraphs library (Bromberger et al., 2017) with algorithms modified as above for computation. The program was then executed on Google Cloud Platform's *N1-standard-16* instance. Network size and time taken for computation are listed in **Table 1.** .

**Table 1.** Graph size after data cleansing and computation time for each city.

| Placename | Process Time | N. Edges | N. Vertices |
|---|---|---|---|
| Barcelona | 38 min 25 s | 54,242 | 37,677 |
| Bari | 5 min 4 s | 25,773 | 19,001 |
| Bilbao | 4 min 11 s | 19,476 | 14,732 |
| Bologna | 8 min 58 s | 28,265 | 21,712 |
| Bordeaux | 31 min 7 s | 43,819 | 32,902 |
| Catania | 12 min 55 s | 27,950 | 21,188 |
| Florence | 9 min 43 s | 29,527 | 22,586 |
| Genoa | 12 min 2 s | 29,940 | 25,173 |
| Las Palmas de Gran Canaria | 5 min 21 s | 25,877 | 19,998 |
| Lille | 29 min 5 s | 39,277 | 30,600 |
| Lyon | 36 min 52 s | 48,481 | 37,390 |
| Malaga | 27 min 14 s | 42,669 | 32,390 |
| Marseille | 39 min 35 s | 49,144 | 39,423 |
| Milan | 1 h 23 min 17 s | 78,356 | 58,450 |
| Montpellier | 36 min 60 s | 40,527 | 31,587 |
| Murcia | 35 min 25 s | 67,282 | 48,009 |
| Nantes | 40 min 33 s | 49,303 | 38,041 |
| Naples | 16 min 19 s | 36,314 | 27,921 |
| Nice | 4 min 19 s | 19,732 | 16,002 |
| Palermo | 5 min 32 s | 21,169 | 15,707 |
| Palma | 6 min 35 s | 24,176 | 17,880 |
| Seville | 23 min 45 s | 45,112 | 31,285 |
| Strasbourg | 20 min 3 s | 36,690 | 27,897 |
| Toulouse | 1 h 14 min 43 s | 69,466 | 53,129 |
| Turin | 20 min 40 s | 43,418 | 31,141 |
| Valencia | 40 min 47 s | 54,894 | 37,573 |
| Zaragoza | 16 min 34 s | 39,397 | 28,117 |
| Paris | 2 h 8 min 40 s | 87,341 | 62,464 |
| Rome | 3 h 3 min 9 s | 124,968 | 97,640 |
| Madrid | 4 h 9 min 28 s | 127,560 | 90,973 |

## 2.4 Visualization with "peak colouring" method

In this research, we mainly explore the data by visualising the results. For this purpose, we introduce a new visualisation method based on "centrality peak height" and "peak radius" as follows.

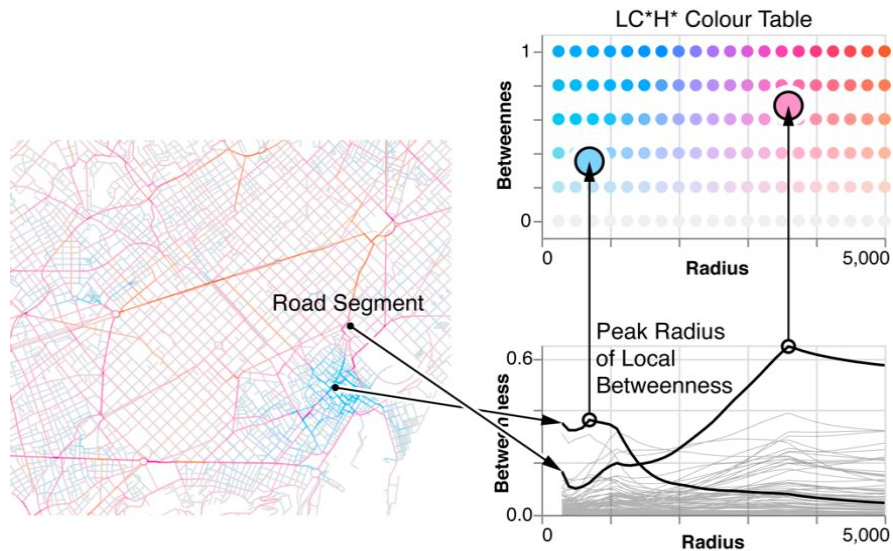

**Fig. 3.** Determining colour for visualisation. Local betweenness is computed within 300-5000 meters with 100 meters step (bottom right). Road segments are coloured according to peak radius of local betweenness and local betweenness at the peak.

In the plot in Fig. 3. bottom right, the horizontal axis is the threshold radius of local betweenness centrality and the vertical axis is the value of local betweenness centrality. Each single line in the bottom right plot represents a feature in the road network.

Each line has a "peak" where the centrality reaches its maximum. Road network features are coloured based on the peak radius and peak height. We used LC*H* colour space (Zeileis et al., 2009) to visualise the vertices. The colour space is made to be perceptually uniform, which means colours with the same L value are perceived as the same lightness, which distinguishes this colour system from others. L corresponds to perceptual lightness and is defined between 0 and 100, C* corresponds to chroma and is defined between 0 and 100, and H* corresponds to hue and is defined between 0 and 360 (degrees).

To colour the edges, betweenness from 0 to 1 are allocated to L from 95 to 60 and C* from 0 to 80, and radius from 300 to 5000 meters are allocated to H* from -135 to 45.

Throughout this research, features with smaller radius peak are visualised in blueish colours and those with larger radius are visualised in reddish colours. In Fig. 3., there are some blueish clusters and reddish lines. Former can be interpreted as road segments important for shorter range transportations while the latter for longer range transportations.

## 2.5 Validating our method

To validate our method, we compare computation result using OSM data with the result using CartoBCN data to test if the algorithm shows similar results and is robust against data quality. Also, we compare our visualisation method with the existing K-means clustering method (Yoshimura et al., 2020) to ensure that they capture similar quality of the road network.

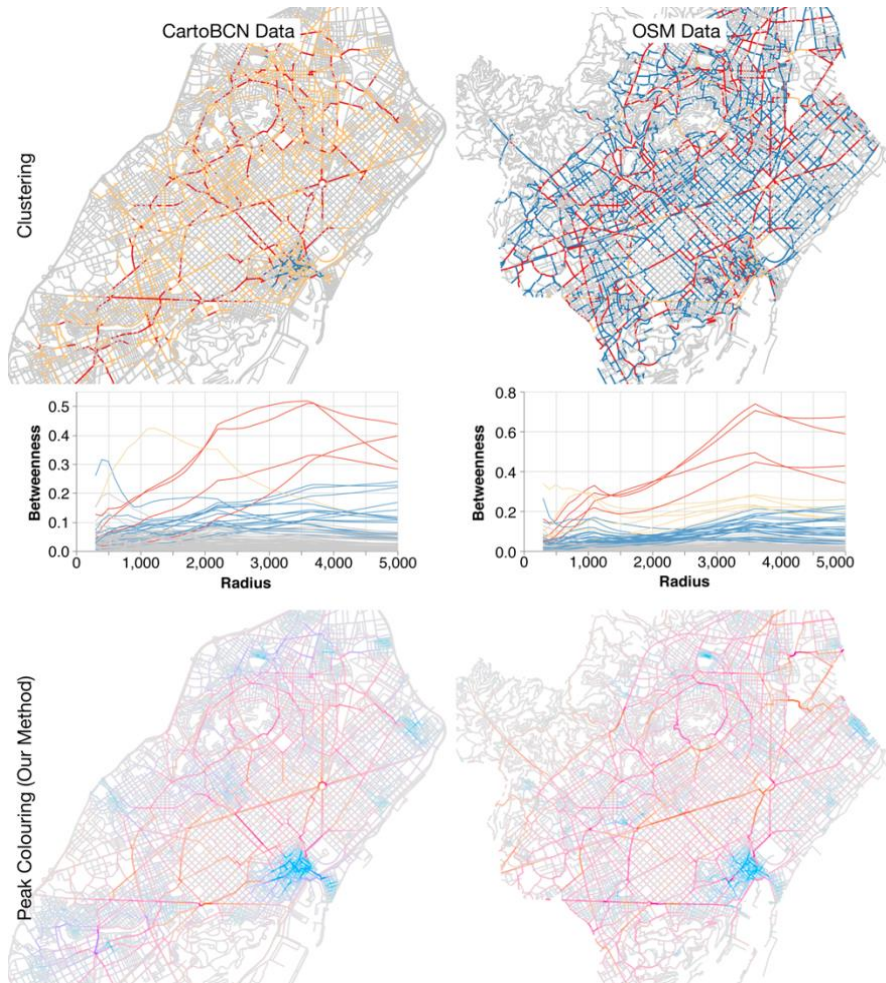

**Fig. 4.** Comparison with reference data/existing method to our method. Top left: reference data with clustering; top right: reference data with our visualisation method; middle: radius-betweenness plot for 100 randomly extracted road segments for the corresponding clustering map; bottom left: OSM data with clustering; bottom right: OSM data with our method. Our colouring method draws out consistent features from analysis results while the clustering method results in different appearances

First we start from replicating Yoshimura et al.(2020)'s K-means clustering method (Yoshimura et al., 2020). In the research, K-means clustering algorithm was used to classify road segment types.

In Fig. 4. top left, a cluster of city-scale interconnect roads (red) and historic centre (blue) are clearly visible. Former cluster consists of road segments whose betweenness centrality increases when analysis radius is increased, while the latter goes the opposite direction. Our peak colouring method is then applied to the same data and it effectively delineates large scale interconnects (reddish) and historic centre (blueish).

However, with OSM data, which is less clean, the result from the clustering algorithm (bottom left) does not match the result for reference data (top left). On the other hand, the result from peak colouring algorithm coincides well with results from both methods for reference data. Each cluster consists of almost the same types of road segments, but the resulting map appears quite different: the clustering algorithm fails to capture the historic centre.

With peak colouring method, the overall appearance of the map has been kept intact. Larger roads appear in reddish colour while the historic centre appears in blueish colour. This shows that the peak colouring method is a robust way to grab a certain quality of road networks.

## 3 Experiments and Analysis

### 3.1 Observation from Visualisations

In this section we will discuss for two most populated cities per country, namely Barcelona, Madrid, Paris, Marseille, Milan and Rome.

Road segments with larger radius peaks had almost the same tendency for all cities. However, those with smaller radius peaks have different structures between cities. Below we will discuss smaller radius peaks for the cities.

In Barcelona, most of the edges with smaller radius peaks are concentrated in a historic district (A). This was almost the only case where the residential area did not strongly stand out on the map. This might be due to its uniformity of the city grid where the southeast half of the analysis area is covered by a 133 metres grid typically seen in *Eixample*. Conversely, there are multiple flocks of edges with smaller radius peaks in Madrid. They are located in the historic district (A), residential area (B: near Los Rosales, C: near Numancia, D: near Arcos, E: near Pinar Del Rey, F: near Peñagrande.)

In Paris, no clear flock of smaller radius peaks can be seen, but clear one is observed in Marseille, Milan, and Rome. For the former, a vague flock appears around Cité island (A), suggesting that Paris has more urban tissues compared to other cities from the viewpoint of pedestrian-scale betweenness analysis. For the latter, in case of Marseille where A is the old port of Marseille, several residential areas are located near C (Saint-André) and road segments in a parking lot of a large shopping centre is picked up. However, this spotting should be regarded as a failure of data cleansing (see B in Marseille). In case of Milan, A is the central station area and B is around the historic centre, and in case of Rome, B is located in an older district while A is a large cemetery. This is also a problem which comes from data quality.

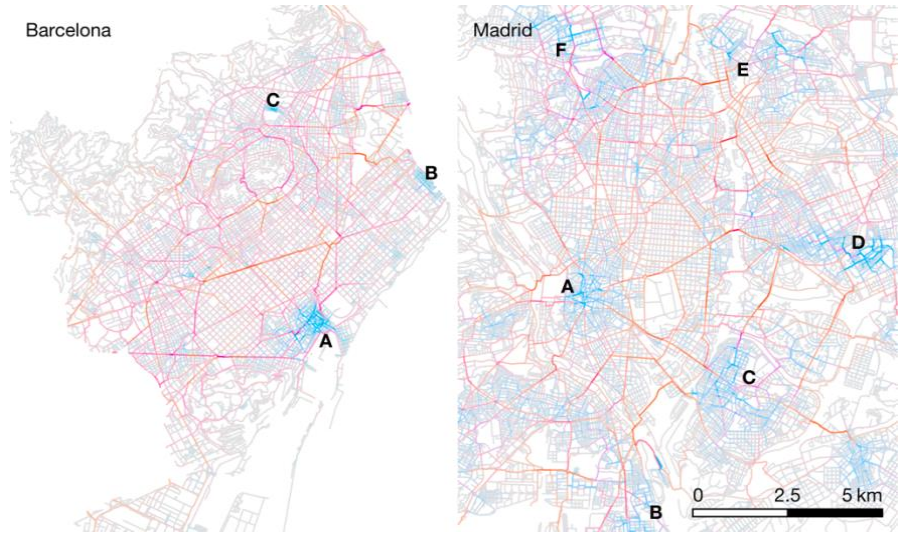

**Fig. 5.** Barcelona (left) and Madrid (right). In Barcelona, most road segments with shorter peak radius are concentrated in historic centre (A). B and C are residential areas. In Madrid, several flocks of short peak radius road segments can be seen. A is a historic centre and B-F are residential areas.

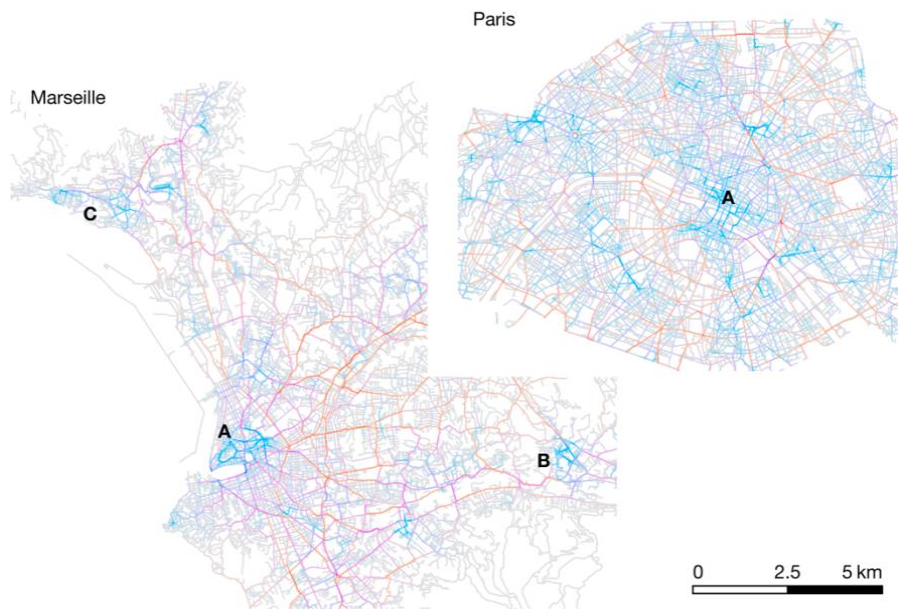

**Fig. 6.** Paris (top right) and Marseille (bottom left). In Paris, short peak radius road segments are relatively distributed compared to other cities. In Marseille, A is a historic centre and C is a residential area. B is road segments near a parking lot.

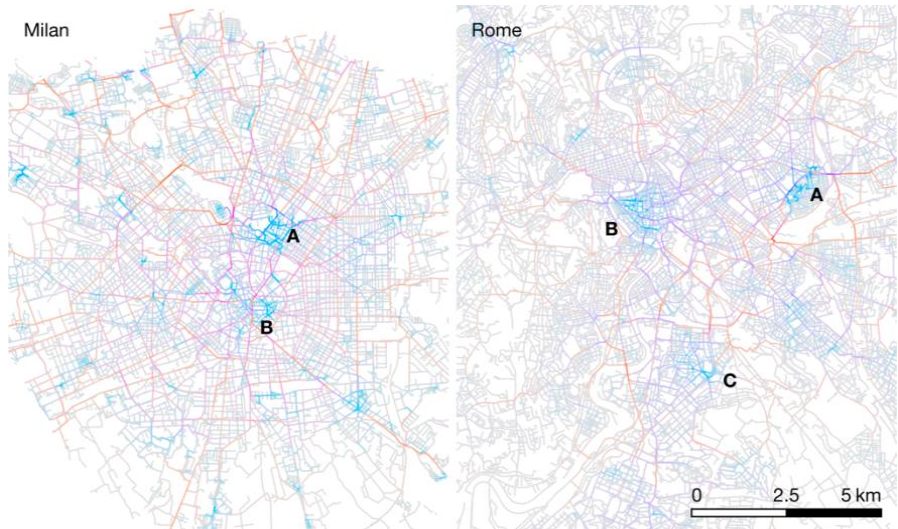

**Fig. 7.** Milan (left) and Rome (right). In Milan, A is the central station area and B is near the historic centre. In Rome, B is located in the historic centre. A is a large cemetery.

### 3.2 Difference Between Small Radius Local Betweenness and Large Radius Local Betweenness

In section 3.1 we have found that road segments which have betweenness peak at smaller radii and those with larger radii have different spatial distribution by observing visualisations. In this section a quantitative analysis is performed to reinforce the observation. The methodology mainly relies on Porta et al. (2009).

First, local betweenness data with radius 300 and 5000 metres are extracted for each city (Fig. 8 left). Former are used as a proxy of the spatial distribution of road segments which have betweenness peaks at smaller radii while latter for those with larger radii. Second, the data is rasterized using 10 metres grids. This step is required to apply smoothing to line features. Third, rasterized data is smoothed using Gaussian kernel (Fig. 8 right). At this step, we have eliminated pixels whose betweenness is lower than 0.2 to focus on important edges: without this step results are too much influenced by less important edges (Fig. 9). Finally, we compute the correlation coefficient between these two data with kernel bandwidth ranging from 100 to 2000 metres.

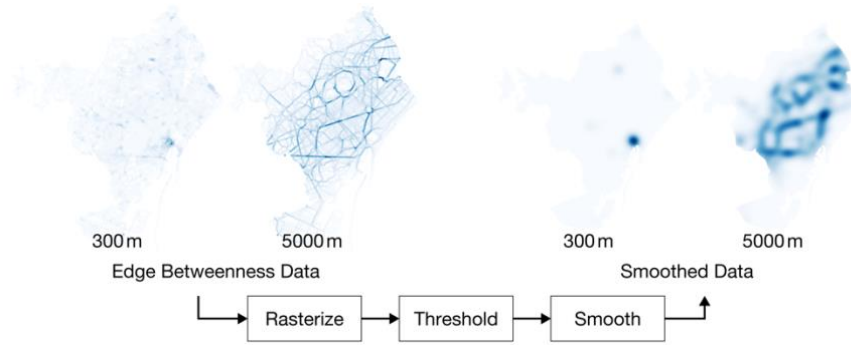

**Fig. 8.** Steps of Correlation analysis. Betweenness data for 300 and 5000 metres (left), rasterized and smoothed (right). Darker colour corresponds to high betweenness.

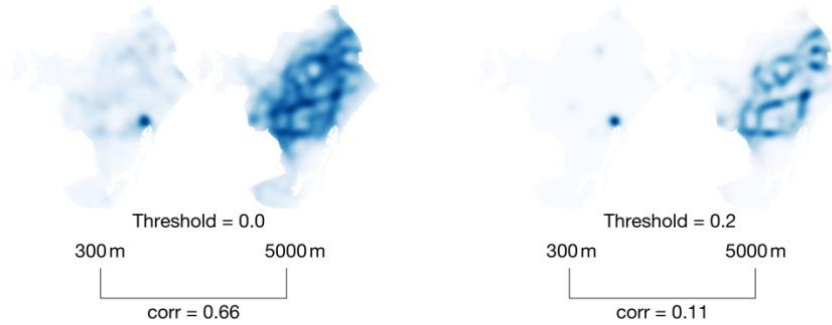

**Fig. 9.** Correlation analysis without threshold (left) and with threshold 0.2 (right). Without thresholding high correlation coefficient is calculated due to relatively unimportant (low betweenness) road segments. By filtering out these road segments one can focus on spatial distribution of important road segments.

The result shows that, with smoothing kernel bandwidth less than 300 metres (pedestrian scale), most cities show correlation coefficient less than 0.4. On the other hand, most cities show correlation coefficient higher than 0.4 with kernel bandwidth larger than 1000 metres (automotive scale). This means that the difference observed above may be important for pedestrians while may be trivial to automotive transportations. See Fig. 10 for the plots.

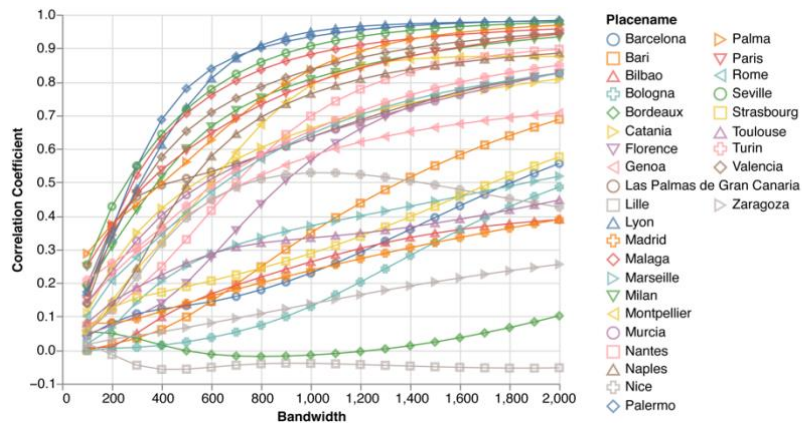

**Fig. 10.** Relationship between smoothing kernel bandwidth and correlation coefficient between local betweenness with radius 300 metres and 5000 metres.

### 3.3 Relationship Between Peak Radius and Road Class

The tendency of distribution of peak radius can also be confirmed by the OSM road class data. Data obtained from the OSM have highway property for each road segment, which roughly indicates road segments' importance mainly for motorized transportation. For that reason, road segments with higher road class are expected to have peaks in larger radius.

To test this, we split our data into two groups for each city. One group is "important" ones which includes highway, trunk, primary, secondary, tertiary, and their links; these tags represent "important roads in a country's system" as noted in the OSM wiki (*OpenStreetMap Wiki*, n.d.). The other group includes the rest such as unclassified or residential roads for example. Fig. 11. shows road segments with importance; darker colour means that the segments are more important. The map lacks highways because we have excluded them as mentioned above, as it is not accessible by pedestrians or bicycles.

Then we count the number of the data per peak radius by importance. In the example plotted in Fig. 12. , road segments without importance include road segments with smaller radius peaks more frequently compared to the other group.

Fig. 13. is the difference of the percentage of peak radius by importance for all 30 cities. A similar tendency can be observed for all the cities analysed: road segments with smaller peak radius are more frequently seen in less important road segments, while those with larger peak radius are located on important road segments. This result comes in accordance with the observation discussed in the previous section.

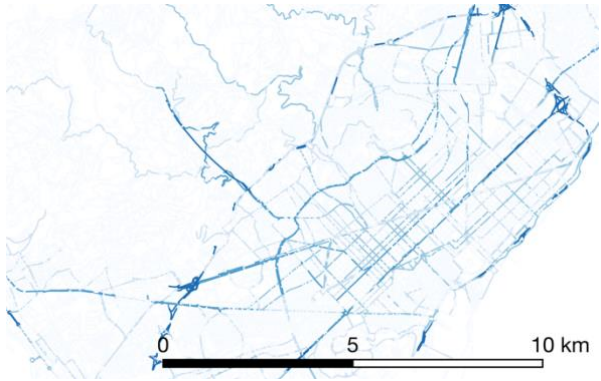

**Fig. 11.** Example of road classes of Barcelona city centre (data source: OSM). Darker colour corresponds to more important roads. Highway is missing because we have only included road segments accessible for pedestrians.

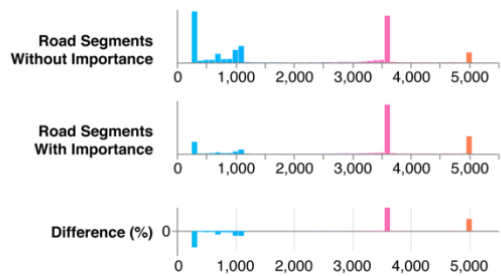

**Fig. 12.** Example of count of peak radius by importance. Top: road segments without importance, Middle: with importance, Bottom: difference.

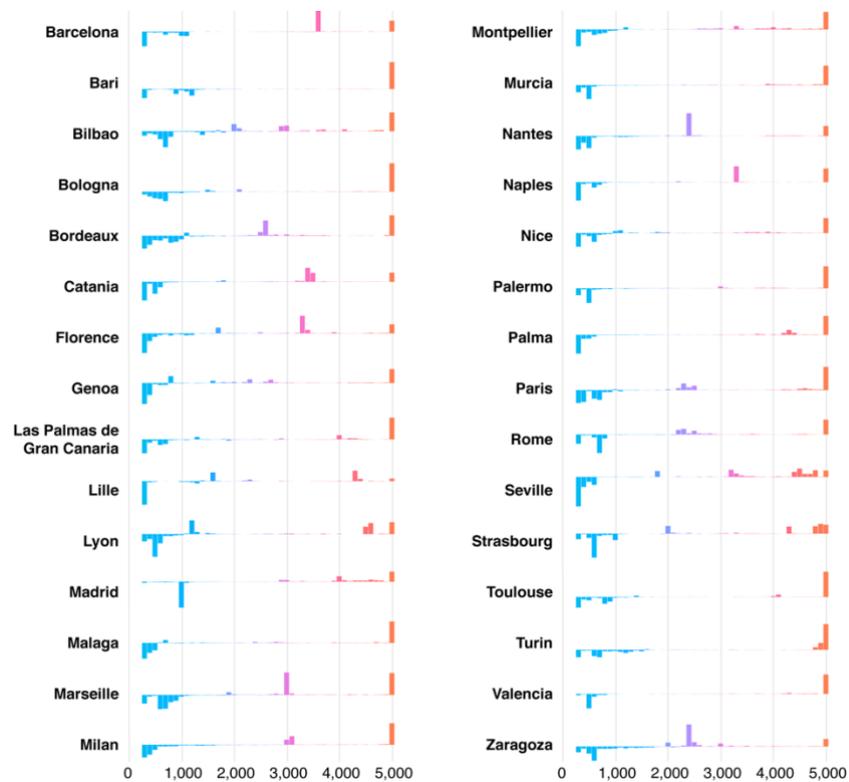

**Fig. 13.** Difference of percentage of peak radius by importance. Positive value means that road segments with the peak radius can be seen more frequently in the road segment with importance.

## 4 Discussion

We analysed 30 European cities using local betweenness analysis and developed its visualisation method. Results show that the indicator successfully measures and uncovers the hidden patterns of road segments that are important for long-range or short-range transportation. Also, we showed that all the cities have a common pattern in their large-scale structure: road segments that are important for longer range activities, are concentrated along larger streets. However, a pattern in smaller scale structure, namely below around 1000 metres of radius, is less common over cities. For example, in Barcelona one can see clear spots of road segments that are important for short-range transportation while in Paris the structure is not as clear.

The results suggest that one needs to focus on features important for pedestrians when comparing road network structures. This is because the main difference between cities lies in the spatial configuration of the road segments which have limited betweenness peaks in smaller radii.

Also, our method is meaningful for city planners and decision-makers. Our method draws characteristics of road segments in a simple but robust way on a single map. This opens up its uses as a consensus building tool where explicability and robustness are needed.

Existing research performs comparative studies on betweenness centrality and urban activity data, such as microscale economic activity data (Yoshimura et al., 2020) or pedestrian flows and urban tissues (Pont et al., 2019). Although this aspect is absent in our research, we provide a comparison among cities through a simple yet powerful visualisation method, resulting from larger dataset than those in the previous studies. The comparison allowed us to grasp morphological patterns in urban street networks, which have not been explicitly pointed out.

Even though our analysis and proposed methodologies provide valuable insights, there are some limitations. First, all vertices and edges are treated equally, but not all roads are the same in reality: they have different traffic capacities, population along roads, amount of greenery, and other features. In future studies, it is necessary to determine if these properties should be taken into account. Second, the phenomena are not well understood in terms of mathematical theory, although we have shown that there are some common properties with exploratory analysis. Further simulation or theoretical research are open to question.

## 5 Conclusion

In this research, we examined the local betweenness centrality measure on 30 European cities to see if the measure is useful for understanding the characteristics of road segments. Our contribution in this paper is as follows:

1. We have shown that spatial distributions of road segments important for long-range transportation and for short-range transportation are largely different at smaller scale while the difference disappears in most cities at larger scale. This means that local betweenness analysis with smaller radius shows different aspect of the cities compared to existing naïve betweenness analysis.

2. Our analysis uncovered that street networks have similar structures among European cities from the viewpoint of long-range transportations while being different from the viewpoint of short-range transportations. This means that analysis focusing on pedestrian scale structure is needed when analysing differences among cities.

3. This indicator allows us to spot places in the city where there is a road network potentially useful for pedestrians. This can be useful for urban re-development. Additionally, its good interpretability makes it useful as a consensus-building tool.

4. Our visualisation method for local betweenness is simple, robust and powerful. The visualisation method can also be applied to other local indicators, thus useful for comparison between other multidimensional local indicators.

These findings can aid urban planners and city authorities in their efforts to develop pedestrian areas, revitalize deteriorated districts or re-habilitate a neighbourhood. Understanding the potential streets segments in a whole network enables us to identify the highest potential and their geographical locations spatially. Thus, the city planner may optimize the infrastructures and the locations of the retail shops to make the district more attractive and alive by increasing the number of pedestrians.